\documentstyle[cizgi,twoside,paper,12pt]{article}

\renewcommand{\theequation}{\the{section}.\arabic{equation}}
\renewcommand{\theequation}{\arabic{section}.\arabic{equation}}
\newcommand{\ben}{\begin{enumerate}}
\newcommand{\een}{\end{enumerate}}
\newcommand{\be}{\begin{equation}}
\newcommand{\ee}{\end{equation}}
\newcommand{\bea}{\begin{eqnarray}}
\newcommand{\eea}{\end{eqnarray}}
\newcommand{\bc}{\begin{center}}
\newcommand{\ec}{\end{center}}
\newcommand{\vs}{\vspace}

\begin{document}
\setcounter{page}{1}

\bc
{\large {\bf
PROLONGATION ALGEBRA AND B\"{A}CKLUND TRANSFORMATIONS OF
DRINFELD-SOKOLOV SYSTEM OF EQUATIONS}}
\ec

\bc
{\em
Ay\c{s}e Karasu(Kalkanl{\i}) \\ and \\ \'{I}smet Yurdu\c{s}en\\
Department of Physics \\
Middle East Technical University \\
06531 Ankara, Turkey}
\ec

\bc
{\em
e-mail:akarasu@metu.edu.tr}
\ec
\vs{2cm}

\bc
{\bf Abstract}
\ec
We show that the Drinfeld-Sokolov
system of equations has a non-trivial prolongation structure. The
closure process for prolongation algebra gives rise to the
$sl(4,c)$ algebra which is used to derive the scattering problem
for the system of equations under consideration.The non-trivial
new B\"{a}cklund transformations and some explicit solutions are given.

\newpage

\baslik{1. Introduction}
\setcounter{section}{1}
\setcounter{equation}{0}
The systems of  nonlinear partial differential equations are
encountered in fundamental particle physics, plasma and fluid
dynamics, statistical mechanics, many areas of solid state
physics, protein dynamics, laser and fiber optics[1].
These equations are usually solved by the use of approximation
techniques, but the range of applicability and usefulness of
these solutions increase the interest on the exact solutions and
methods of testing for complete integrability. During the past
three decades the developments in mathematical physics show
that the completely integrable systems of nonlinear partial
differential equations have very rich mathematical structures
such as the existence of Lax pair, bi-Hamiltonian structures,
recursion operators and applicability of inverse scattering
methods. One of the effective methods to test integrability
is the prolongation structure technique of Wahlquist
and Estabrook[2]. This geometrical method immediately proceeds
with an attempt to construct the linear spectral problem or
another device, such as B\"{a}cklund transformations. If we
are successful we have not only tested our system of equations
for complete integrability but also constructed a device with
which to integrate the system. Dodd and Fordy [3],[4] made the
method more algorithmic and put it in an algebraic, instead of
differential geometric framework. The first step is to start with a
differential equation and derive a set of generators and relations
for an incomplete Lie algebra. The second step is to complete this
Lie algebra and find a finite matrix representation for the
derived set of generators. Then, one expects that B\"{a}cklund
transformations, which are relations among solutions of the nonlinear
differential  equations under consideration, may be  obtained.
For a review the reader is referred to an article by Harrison[5] and
references quoted therein.\\
The integrable systems appear not one at
a time, but in big families which are called hierarchies. First,
Korteweg-de Vries(KdV)-hierarchy was invented and then infinitely
many generalized KdV-hierarchies were found. They were unified to
a single one large Kadomtsev- Petviashvili(KP)-hierarchy[6].
Very recently, it is shown by G\"{u}rses
and Karasu [7] that the system of equations \bea u_t &=&
-u_{xxx}+6uu_x+6v_x \nonumber\\ v_t &=& 2v_{xxx}-6uv_x \eea admits
a recursion operator and a bi-Hamiltonian structure,therefore it
has constants of motion. The system (1.1) belongs to a type of
equations which are called "quasi-polynomial flows". The Lax pair
for this system was first given by Drinfeld and Sokolov [8] and
later by Bogoyavlenskii [9]. Under the scale transformations,
this system of
equations reduces to a special case of the KP-hierarchy which was
given by Satsuma and Hirota [10]. They also gave the one-soliton
solutions. Recently, auto-Backlund transformations and certain
analytical solutions are obtained by Tian and Gao [11] via Painl\'{e}ve
analysis. In this work, we used the prolongation method to derive the
linear scattering problem for the system (1.1). We obtained B\"{a}cklund
transformations by using pseudopotentials.
\baslik{2. Prolongation Structure}
\setcounter{section}{2}
\setcounter{equation}{0}
By introducing the variables $p = u_x$, $q = v_x$, $r = p_x$, $s = q_x$,
the system of equations (1.1) can be represented by the set of
2-forms,
\bea \alpha_1 & = & du \wedge dt - p \; dx \wedge dt \nonumber \\
\alpha_2 & = & dp \wedge dt - r \; dx \wedge dt \nonumber \\
\alpha_3 & = & dv \wedge dt - q \; dx \wedge dt \nonumber \\
\alpha_4 & = & dq \wedge dt - s \; dx \wedge dt \nonumber \\
\alpha_5 & = & du \wedge dx - dr \wedge dt +6 (up+q) \; dx \wedge
dt \nonumber \\ \alpha_6 & = & dv \wedge dx + 2 ds \wedge dt - 6uq
\; dx \wedge dt \eea which constitutes a closed ideal $I$,such
that $dI \subset I$.\\
We extend the ideal $I$
by adding to it the system of 1-forms,
\be
w^A = dy^A + F^A \; dx + G^A \; dt, \qquad  A =1,  \ldots , N
\ee
where $F^A$ and $G^A$ are functions of $(u,v,p,q,r,s,y^A)$,
which are assumed in the form $F^A = F^A_B y^B $,
$G^A = G^A_B y^B $.
The extended ideal must be closed under exterior differentiations.
This requirement gives the set of partial differential equations
for $F^A$ and $G^A$. Dropping the indices for simplicity, we have
\bea
F_p = F_q = F_r = F_s = 0 ,\qquad
G_s = 2 F_v  ,\qquad
G_r =- F_u , \nonumber \\
{}pG_u + qG_v +rG_p + sG_q - 6(up+q) F_u + 6uq F_v - [F,G] =0
\eea
where
\be
[F, G] =F^B \frac{\partial G}{\partial y^B} -
G^B \frac{\partial F}{\partial y^B}. \nonumber\\
\ee
Next, we integrate equations (2.3). In reaching the results
we equate the coefficients of quadratic terms in F to zero, since these
coefficients are in the centre of prolongation algebra[4].
The final result is,
\bea
F & = & X_1 + X_2 u + X_3 v \nonumber \\
G & = & X_0 + (-r+3u^2+6v) X_2 + 2 (s-3uv) X_3
- pX_4 - uX_5 \nonumber \\
& & - \frac{u^2}{2} X_6 + 2q X_7 + 2v X_8 + v^2 X_9 + 2uv X_{10}
\eea
where $X_0,X_1,X_2,X_3$ are constants of integration, depending
on $y^A$ only. The remaining elements are,
\bea
X_4  =  [X_1,X_2],\qquad
X_5  =  [X_1,X_4], \qquad
X_6  =  [X_2,X_4], \nonumber\\
X_7  =  [X_1,X_3], \qquad
X_8  =  [X_1,X_7], \qquad
X_9  =  [X_3,X_7], \nonumber\\
X_{10}  =  [X_2,X_7] .
\eea
The integrability conditions  impose following restrictions on
$X_i (i =1,  \ldots , 10)$,
\bea
[X_2,X_3] = 0 , \qquad
[X_2,X_6] = 0 , \qquad
[X_3,X_9] = 0 , \qquad
[X_1,X_0] = 0 , \nonumber\\
{}[X_3,X_4] = 6X_3-2X_{10}, \qquad
[X_2,X_9] + 2 [X_3,X_{10}] = 0 , \nonumber\\
{}[X_2,X_{10}] - \frac{1}{4} [X_3,X_6] = 0, \nonumber\\
{}[X_1,X_6] + 2[X_2,X_5] - 6X_4 = 0 ,\nonumber\\
{}[X_1,X_9] + 2[X_3,X_8] = 0, \nonumber\\
{}[X_1,X_5] - [X_2,X_0] = 0, \nonumber\\
{}2[X_1,X_8] + [X_3,X_0] + 6X_4 = 0 \nonumber\\
{}[X_1,X_{10}] + [X_2,X_8] - \frac{1}{2} [X_3,X_5] - 3X_7 = 0.
\eea
Using the Jacobi identities we obtain further relations:
\bea
[X_1,X_9] = [X_3,X_8] = 0, \qquad
[X_2,X_9] = [X_3,X_{10}] = 0, \nonumber\\
{}[X_2,X_{10}] = [X_3,X_6] = 0,\qquad
[X_2,X_7] = [X_3,X_4] \nonumber\\
{}[X_1,X_{10}] - [X_4,X_7] - [X_2,X_8] = 0, \nonumber\\
{}[X_1,X_6] = [X_2,X_5] = 2X_4,\qquad
[X_4,X_6] = -2X_6,  \nonumber\\
{}[X_3,X_5] + [X_7,X_4] + 2[X_1,X_{10}] = 6X_7.
\eea
In order to find the Lie algebra generated by F and matrix
representations of the generators $\{X_i\}_0^{10}$, we follow
the strategy of Dodd-Fordy[4]. It can be summarized as follows:\\
1. Locate elements of the center of the algebra. Assuming
the algebra to be (semi-)simple equate these elements with zero.\\
2. Locate a nilpotent and semi-simple element.\\
3. Embed these elements in a simple Lie algebra $g$.\\
4. Express the remaining generators of the prolongation
algebra as linear combinations of a suitable basis of $g$.\\
5. Use the fundamental representation of $g$ to
generate a linear scattering problem.\\
First we reduce the number of elements. By using equations (2.6),
(2.7) and (2.8) we get
\be
X_9 = 0, \qquad
X_6 = 2X_2 ,\qquad
X_{10} = 2X_3.
\ee
Next, we locate nilpotent and neutral elements. Equations (2.6) and
(2.9) give that  $X_2$ is nilpotent and $X_4$ is neutral element.
Let us note that the system of equations in (1.1) has the following
scale symmetry
\be
x \rightarrow \lambda^{-1} x , \qquad
t \rightarrow \lambda^{-3} t , \qquad
u \rightarrow \lambda^{2} u ,\qquad
v \rightarrow \lambda^{4} v , \nonumber
\ee
which implies that the elements $X_{i}$ must satisfy
\bea
X_0 \rightarrow \lambda^{3}X_0 , \qquad
X_1 \rightarrow \lambda X_1 , \qquad
X_2 \rightarrow \lambda^{-1}X_2 ,\qquad
X_3 \rightarrow \lambda^{-3} X_3 , \nonumber \\
{}X_4 \rightarrow  X_4 ,\qquad
X_5 \rightarrow \lambda X_5 ,\qquad
X_7 \rightarrow \lambda^{-2} X_7 ,\qquad
X_8 \rightarrow \lambda^{-1} X_8.
\eea
By using the basis elements,
we try to embed the prolongation algebra into  $sl(n+1,c)$. Starting
from the cases $n=1,2$, we found that $sl(2,c)$ and $sl(3,c)$ can not be
the whole algebra. The simplest non-trivial closure is in terms of
$sl(4,c)$. Without giving the details here we present the results:
\bea
X_1  = \left( \begin{array}{cccc}
0 & 1 & 0 & 0 \\
0 & 0 & 2 & 0 \\
0 & 0 & 0 & 1 \\
\frac{\lambda^4}{2} & 0 & 0 & 0
\end{array} \right),  \qquad
X_2  = \left( \begin{array}{cccc}
0 & 0 & 0 & 0 \\
1 & 0 & 0 & 0 \\
0 & 0 & 0 & 0 \\
0 & 0 & 1 & 0
\end{array} \right) , \nonumber \\
{}X_3  = \left( \begin{array}{cccc}
0 & 0 & 0 & 0 \\
0 & 0 & 0 & 0 \\
0 & 0 & 0 & 0 \\
-\frac{1}{2} & 0 & 0 & 0
\end{array} \right) , \qquad
X_4  = \left( \begin{array}{cccc}
1 & 0 & 0 & 0 \\
0 & -1 & 0 & 0 \\
0 & 0 & 1 & 0 \\
0 & 0 & 0 & -1
\end{array} \right) , \nonumber \\
{}X_5  = \left( \begin{array}{cccc}
0 & -2 & 0 & 0 \\
0 & 0 & 4 & 0 \\
0 & 0 & 0 & -2 \\
\lambda^4 & 0 & 0 & 0
\end{array} \right),  \qquad
X_7  = \left( \begin{array}{cccc}
0 & 0 & 0 & 0 \\
0 & 0 & 0 & 0 \\
-\frac{1}{2} & 0 & 0 & 0 \\
0 & \frac{1}{2} & 0 & 0
\end{array} \right) , \nonumber \\
{}X_8  = \left( \begin{array}{cccc}
0 & 0 & 0 & 0 \\
-1 & 0 & 0 & 0 \\
0 & 1 & 0 & 0 \\
0 & 0 & -1 & 0
\end{array} \right),\qquad
X_0  = \left( \begin{array}{cccc}
0 & 0 & 0 & -8  \\
-4\lambda^4 & 0 & 0 & 0 \\
0 & -2\lambda^4 & 0 & 0 \\
0 & 0 & -4\lambda^4 & 0
\end{array} \right),
\eea

These generators satisfy the following commutation
relations:
\bea
[X_2,X_4] = 2X_2 ,\qquad
[X_2,X_3] = 0,\qquad
[X_3,X_4] = 2X_3 ,\nonumber \\
{}[X_1,X_2] =X_4 ,\qquad
[X_1,X_3] = X_7 ,\qquad
[X_4,X_7] = 0, \nonumber \\
{}[X_3,X_7] = 0 ,\qquad
[X_1,X_7] = X_8 , \qquad
[X_3,X_8] = 0 , \nonumber \\
{}[X_2,X_7] = 2X_3 ,\qquad
[X_1,X_4] = X_5 ,\qquad
[X_2,X_5] = 2X_4 , \nonumber \\
{}[X_3,X_5] = 2X_7 ,\qquad
[X_2,X_8] = 2X_7 ,\qquad
[X_0,X_1] = 0 , \nonumber \\
{}[X_2,X_0] = [X_1,X_5] , \qquad
[X_3,X_0] = -6X_4 -2[X_1,X_8]
\eea
By substituting the matrix representations of the generators into
equation (2.5) we form the matrices F and G. Using equations
$y_x = -Fy , y_t = -Gy$ ,the linear
scattering problem can be obtained as,
\be
\left( \begin{array}{c}
y^1 \\
y^2 \\
y^3 \\
y^4
\end{array} \right)_x =
\left( \begin{array}{cccc}
0 & -u & 0 & -\frac{\lambda^4}{2}+\frac{v}{2} \\
-1 & 0 & 0 & 0 \\
0 & -2 & 0 & -u \\
0 & 0 & -1 & 0 \\
\end{array} \right)
\left( \begin{array}{c}
y^1 \\
y^2 \\
y^3 \\
y^4
\end{array} \right)
\ee,
\be
\left( \begin{array}{c}
y^1 \\
y^2 \\
y^3 \\
y^4
\end{array} \right)_t =
\left( \begin{array}{cccc}
p & 4\lambda^4-4v+r-2u^2 & q & \lambda^4 u+s-uv \\
-2u & -p & 2\lambda^4-2v & -q \\
0 & 4u & p & 4\lambda^4-4v+r-2u^2 \\
8 & 0 & -2u & -p \\
\end{array} \right)
\left( \begin{array}{c}
y^1 \\
y^2 \\
y^3 \\
y^4
\end{array} \right)
\ee \\which is equivalent to the scalar Lax equation
\be
L\psi = (\partial^4 - 2u\partial^2 - 2u_x\partial -
u_{xx} + u^2 + v )\psi = \lambda^4\psi
\ee
where $\psi = y^4$ and $\lambda = constant$. The corresponding time
evolution of $\psi$ is
\be
\psi_t = (-4\partial^3 + 6u\partial + 3u_x )\psi.
\ee

\baslik{3.B\"{a}cklund Transformations }
\setcounter{section}{3}
\setcounter{equation}{0}
Within the prolongation scheme, B\"{a}cklund transformations
can be derived by assuming the new solution variables as
functions of old ones and the ratios of pseudopotentials[5].
For this purpose ,let us define new variables
\be
\alpha = \frac{y^1}{y^4},\qquad
\beta = \frac{y^2}{y^4},\qquad
\gamma = \frac{y^3}{y^4}.
\ee
By using  equations (2.14) and (2.15) we can find the equations
satisfied by $\alpha$,$\beta$ and $\gamma$,
\bea
\alpha_x &=& \alpha \gamma - u\beta +\frac{1}{2}(v-\lambda^4),\nonumber\\
\beta_x &=& -\alpha + \beta \gamma, \nonumber\\
\gamma_x &=& \gamma^2 - 2\beta - u, \nonumber\\
\alpha_t &=& -8\alpha^2+2(p+2u\gamma)\alpha+(4\lambda^4-4v+r-2u^2)\beta+
          q\gamma + (\lambda^4 u + s - uv), \nonumber\\
\beta_t &=& -2u\alpha + 2(u\gamma - 4\alpha)\beta +2(\lambda^4-v)\gamma-
          q , \nonumber\\
\gamma_t &=& 2u\gamma^2+2(p-4\alpha)\gamma + 4u\beta +
          (4\lambda^4-4v+r-2u^2).
\eea
The compatibility conditions $\alpha_{xt}=\alpha_{tx}$ and
$\gamma_{xt}=\gamma_{tx}$ hold if $u$ and $v$ satisfy equation (1.1),
while $\beta_{xt}=\beta_{tx}$ holds automatically. One can easily
check that the function $\beta$ satisfies the following equation,
\be
\beta_t-2\beta_{xxx}+6(u+2\gamma_x)\beta_x=0.
\ee
This means that
\bea
\tilde{u} &=& u + 2\gamma_x ,\qquad \nonumber\\
\tilde{v} &=& c_1\beta + c_2
\eea
are the new solutions of equations (1.1) if
\be
(4\beta^2 +c_1 \beta)_x=0
\ee
where $c_1$ and $c_2$ are constants.
We  note that the same results for   $\tilde{u}$ and $\tilde{v}$ were
obtained  when we followed the step(3) of ref.[5].
By seting $y_4 = \psi$ and using  equations (2.14),(2.15) and (3.1) we
obtain
\bea
\tilde{u} &=& u - 2 \frac{\psi_{xx}}{\psi} +
               2 \frac{\psi_x^2}{\psi^2} ,\nonumber \\
\tilde{v} &=& c_1(-\frac{u}{2} + \frac{\psi_{xx}}{\psi}) + c_2.
\eea
Here, $u$ is a known solution of equations (1.1) and $\psi$ is the
solution of equations (2.16) ,(2.17) satisfying the condition
\be
\{\frac{1}{2\psi^2}[\psi_{xx}-u\psi][2(\psi_{xx}-u\psi)+c_1\psi]\}_x=0.
\ee
which is equivalent to (3.5).\\
Next, we consider the simple case $u = v =0$ as the known solution of
(1.1) and find a new solution. With this choice, we find that
\be
\psi = d_1 e^{-\lambda(4\lambda^2t-x)} + d_2 e^{\lambda(4\lambda^2t-x)}
  + d_3 e^{i\lambda(4\lambda^2t+x)} + d_4 e^{-i\lambda(4\lambda^2t+x)}
\ee
is a solution of equations (2.16) and (2.17) where $d_1,d_2,d_3$ and
$d_4$ are constants. Substituting (3.8) into condition (3.7)
we obtain two sets of solutions for $\psi$ whereas $\{d_1 = d_2 = 0,
c_1 = 4\lambda^2\}$ and \\$\{d_3 = d_4 = 0, c_1 = -4\lambda^2\}$.\\
The respective solutions for $\tilde{u}$ are
\bea
\tilde{u_1} &=& \frac{8d_3d_4\lambda^2e^{2i\lambda(4\lambda^2t+x)}}
                 {[d_4 + d_3e^{2i\lambda(4\lambda^2t+x)}]^2},
                 \nonumber \\
\tilde{u_2} &=& \frac{-8d_1d_2\lambda^2e^{2\lambda(4\lambda^2t-x)}}
                 {[d_1 + d_2e^{2\lambda(4\lambda^2t-x)}]^2}.
\eea
where in both cases $\tilde{v} = c_2 - 2\lambda^4 = constant$.
Thus, starting from trivial background, we obtained
the one soliton solution of KdV equation with $\tilde{v} =
constant$.\\
In order to find the more general B\"{a}cklund transformations
for system (1.1), we assume that new solutions $\tilde{U}$
and $\tilde{V}$ are functions of old variables $u,v,p,q,r,s$
and $\alpha,\beta,\gamma$ which are the ratios of prolongation
variables satisfying the equations (3.2). After some straightforward
but long calculations we obtained the following results:
\bea
\tilde{U} &=& \frac{\Omega}{2}-\Delta-\frac{2}{\gamma}(\alpha
               +2\beta\gamma) ,\qquad \nonumber\\
\tilde{V} &=& \Theta_{xx} -\frac{\Delta}{\gamma}\Theta_x -\Theta^2,
\eea
where
\bea
\Omega &=& -\frac{1}{\gamma^2}(p\gamma+u^2+4\beta u +4\beta^2
              -4\alpha\gamma) ,\qquad \nonumber\\
\Delta &=& 2\beta-2\gamma^2+u,\qquad \nonumber\\
\Theta &=& \frac{1}{4\gamma^2}[-u^2+2(\gamma^2-2\beta)u-4(\alpha
         \gamma+\beta^2)],
\eea
and $\Theta$ must satisfy the condition
\be
\Theta_t = 2\Theta_{xxx}+3\Omega\Theta_x.
\ee
Thus, if any solutions $u$ and $v$ to Drinfeld-Sokolov system
of equations are known and if $\alpha,\beta,\gamma$ are solutions
of (3.2) satisfying the condition (3.12),
then  $\tilde{U}$ and $\tilde{V}$ are the
new solutions of (1.1).\\
If we take the trivial seed solutions $u = v =0$, we get
\bea
\tilde{U} &=& -\frac{2\beta^2}{\gamma^2}+2(\gamma^2-2\beta),
               \qquad \nonumber\\
\tilde{V} &=& -\frac{1}{\gamma^4}(
               17\beta^4+8\beta^3\gamma^2-10\beta^2\alpha\gamma
               +\alpha^2\gamma^2)+(\frac{\lambda^4}{2}+\frac{c_0}{18}),
\eea
with the condition
\be
\beta[-56\beta^4+48\beta^2\gamma\alpha-3\beta\gamma^2\lambda^4
     -8\gamma^2\alpha^2+\gamma^4\lambda^4]=0.
\ee
By setting $y^4=\psi$, one can write these expressions in terms
of the solutions of Lax equations (2.16),(2.17) and see that our
results are more general than the given by Tian and Gao[11].\\
As an example, we consider the simple case,$u = v =0$ as the known
solutions of (1.1) and we obtain the following two sets of explicit
solutions:
\be
\tilde{U}=\frac{2c_{1}^{2}}{(c_1 x + c_2)^2},\qquad
\tilde{V}=\frac{c_0}{18}.
\ee
and
\bea
\tilde{U} &=&\frac{30(c_1+x^2)\{4x[5c_1[2(24t-x^3)-3c_1x]-(c_2+3x^5)]
              +15(x^2+c_1)^3]\}}{\{5c_1[2(24t-x^3)-3c_1x]-(c_2+3x^5)
              \}^2},
\nonumber\\
\tilde{V} &=&-\frac{120x[c_2+3x^5+15c_1^2x-10c_1(24t-x^3]^3}
             {\{5c_1[2(24t-x^3)-3c_1x]-(c_2+3x^5)\}^4},
\eea
where $c_0,c_1,c_2$ are constants.\\
As we observed that there are no solitary wave solutions belonging to
the class (3.13). Very recently, all the special solutions of (1.1)
are obtained by Karasu and Sakovich [12].

\baslik{4. Conclusion}

In this work we have rederived the linear scattering problem for
Drinfeld-Sokolov system of equations by using the prolongation
algorithm. We found the auto-B\"{a}cklund transformations and
some exact solutions of these equations.
The system can be integrated by the method of
inverse scattering problem associated with the fourth order Lax
operator L, which was developed by Iwasaki [13]. It is known that
the most general B\"{a}cklund transformation would be the one
which utilizes the infinite dimensional algebra and not all of
finite algebras give rise to  B\"{a}cklund transformations.
Without searching if the incomplete algebra  given in (2.7) is
finite or infinite dimensional we used a finite dimensional
representation of prolongation algebra and derived  non-trivial
B\"{a}cklund transformations. Thus, the methods given in [4] and
[5] are quite useful from the practical point of view, for the
systems of nonlinear partial differential equations. Finally, we
note that a close connection between some stationary flows
associated with fourth-order Lax operators and generalisations
of some integrable Hamiltonian systems with quartic potentials
is known [14]. The equations (2.16) and (2.17) can be considered
in this context.\\ \\ \\

We would like to thank Dr.Atalay Karasu and Dr.Sergei Sakovich
for useful discussions. We also thank to anonymous referee
for valuable comments.
This work is supported in part by
the Scientific and Technical Research Council of Turkey (TUBITAK).

\begin{reference}
\item Fordy,A.P.(ed.),{\it Soliton Theory: A Survey of Results},
 Manchester University Press,(1990).
\item Wahlquist, H.D. and Estabrook, F.B.,
J.Math.Phys.,{\bf 16},1(1975).
\item Dodd,R.,Fordy,A.,Physics Letters, {\bf A 89},
168(1982).
\item Dodd,R.,Fordy,A.,Proc.R.Soc.Lond.,{\bf A 385},
389(1983).
\item Harrison,K.,Journal of Nonlinear Mathematical Physics,{\bf2},
201(1995).
\item Date,E., Jimbo,M., Kashiwara,M., Miwa,T.,
In {\it Non-linear Integrable Systems, Classical and Quantum
Theory}. Proceedings of RIMS Symposium, Singapore,39(1983).
\item G\"{u}rses,M., Karasu,A.,
Physics Letters, {\bf A 251},347(1999).
\item Drinfeld,V.G., Sokolov,V.V.,
{\it Proceedings of S.L. Sobolev Seminar}, Novosibirsk, {\bf 2},
5(1981) (in Russian).
\item Bogoyavlenskii,O.I.,
Russian Math. Surveys ,{\bf 45}, 1(1990).
\item Satsuma,J.,Hirota,R.,
J.Phys.Soc.J., {\bf 51}, 3390(1982).
\item Tian,B., Gao,Y.,Physics Letters, {\bf A 208},
193(1995).
\item Karasu(Kalkanl{\i}),A., Sakovich,S.Yu.,
Preprint arXiv:nlin.SI/0102001(2001).
\item Iwasaki, K., Japan J. Math.{\bf 14 },1(1988).
\item Baker,S., Enolskii,V.Z., Fordy,A.P.,Physics Letters,{\bf A 201},
167(1995).
\end{reference}

\end{document}